\documentclass[aps,floatfix,prd]{revtex4}
\usepackage{graphicx}
\usepackage{dcolumn}
\usepackage{bm}
\usepackage{amsmath}
\begin{document}

\def\dq{\frac{d^3q}{(2\pi)^3}\,} 
\def\be{\begin{equation}}
\def\ee{\end{equation}}
\def\A{{\bf A}}
\def\B{{\bf B}}
\def\x{{\bf x}}
\def\r{{\bf r}}
\def\y{{\bf y}}
\def\k{{\bf k}}
\def\s{{\bf s}}
\def\S{{\bf S}}
\def\l{{\bf l}}
\def\L{{\bf L}}
\def\q{{\bf q}}
\def\z{{\bf z}}
\def\D{{\bf D}}
\def\P{{\bf P}}
\def\p{{\bf p}}
\def\E{{\bf E}}
\newcommand{\bsigma} {{\mbox{\boldmath $\sigma$}}}
\newcommand{\balpha} {{\mbox{\boldmath $\alpha$}}}
\newcommand{\bnabla} {{\mbox{\boldmath $\nabla$}}}
\newcommand{\bdel} {{\mbox{\boldmath $\nabla$}}}
\newcommand{\bPi}{{\mbox{\boldmath $\Pi$}}}

\title{Light Hybrid Meson Mixing and Phenomenology}

\author{E.S. Swanson}
\affiliation{
Department of Physics and Astronomy,
University of Pittsburgh,
Pittsburgh, PA 15260,
USA.}

\date{\today}

\begin{abstract}
A simple constituent model of gluodynamics that is motivated by lattice field theory  and the QCD Hamiltonian in Coulomb gauge is applied to descriptions of hybrid meson flavor mixing and vector hybrid configuration mixing.
Good agreement with lattice gauge computations is obtained for flavor multiplet masses, while mixing angles are in approximate agreement, given large errors. The configuration mixing results are also in rough agreement with lattice NRQCD calculations. Thus the viability of constituent gluon models of hybrid hadrons and glueballs is supported.
The results suggest that a flavor multiplet of vector hybrids should appear with masses of approximately 2100, 2200, and 2300 MeV and that the isovector vector hybrid decay constant is about 20 MeV. Similarly, the $\pi_1$ exotic hybrid should have isospin partner states near 1760 and 1900 MeV, and it is suggested that the recently seen $\eta_1$ hybrid signal is the latter state.
\end{abstract}

\maketitle 

\section{Introduction}

Although the notion of hadrons with explicit gluonic degrees of freedom has been accepted for nearly 50 years, remarkably little is known about these particles. Nevertheless, it is hoped that steady progress in lattice field theory coupled with new effort at the GlueX, PANDA, and BESIII experiments will finally shed light on these enigmatic states.

Models of gluonic degrees of freedom have generally assumed that they comprise collective string-like excitations or some sort of quasiparticle\cite{Meyer:2015eta}. Recently, the quasiparticle picture has received support from lattice field theory, where a measurement of the low lying charmonium hybrid spectrum strongly suggests that an axial quasigluon with an effective mass something less than 1000 MeV can explain the observed systematics\cite{Dudek:2011bn}. 
This intriguing observation has revived interest in constituent gluon models, wherein early work\cite{HM} has evolved into more sophisticated modelling that builds on QCD\cite{Szczepaniak:2001rg,Swanson:1998kx,Guo:2008yz,Szczepaniak:2003mr}. This modelling starts with the QCD Hamiltonian in Coulomb gauge and constructs gluonic quasiparticles with an Ansatz that builds gluonic correlations in the vacuum. It is expected that the resulting field theory admits reliable Fock space truncations which greatly enhances the ability to model and compute hadronic properties.

The purpose of this work is to examine the viability of a simple constituent gluon model of hybrid properties that is based on the considerations just given. This will be done by computing flavor mixing of light hybrid mesons and configuration mixing in vector mesons. The calculation is also of interest because the flavor mixing mechanism is very different from that for canonical mesons since the quark pair is in a color octet state. In particular, the leading order mechanism annihilates and creates quark pairs via coupling to the instantaneous Coulomb gauge interaction, while the next order mechanism couples hybrid mesons to low lying glueballs. In this work, these glueballs are described with the same degrees of freedom and dynamics as employed for hybrids, thereby testing consistency of the model.
Model validation is possible because a comprehensive computation of the light meson spectrum in lattice QCD has been made -- a computation that includes isoscalar and isovector low lying hybrid mesons and their mixing angles\cite{Dudek:2013yja}. 

A  motivation for the constituent gluon model employed here will be given in the next section. Section \ref{sect:3} applies the model to a computation of the light meson spectrum, which fixes parameters for the light hybrid spectrum, which permits investigation of flavor mixing in Section \ref{sect:4}. Comparison to lattice gauge results is made in Section \ref{sect:5}. A simple extension to vector hybrid-vector ($q\bar q$) meson mixing is presented in \ref{sect:6}. A examination of the implications of the results on the light hybrid and meson spectrum is made in Section \ref{sect:7} and we conclude in Section \ref{sect:8}.

\section{Constituent Gluon Model}
\label{sect:2}

\subsection{Model Construction}

A model dynamics capable of describing the interaction of constituent quarks and axial quasigluons has been developed\cite{Szczepaniak:2001rg} and has been applied to glueballs\cite{Szczepaniak:2003mr},
the gluelump and hybrid spectra\cite{Guo:2008yz,Szczepaniak:2006nx,Guo:2007sm}, and to heavy hybrid decays\cite{Farina:2020slb}. We briefly summarize the salient features of the model to place the subsequent development in context.

The starting point for the dynamical model is the Hamiltonian of QCD in Coulomb gauge. This gauge choice is expedient for model building because Gauss's law has been resolved,  all degrees of freedom are physical, and an explicit interaction potential that operates between quarks and gluons emerges. This ``Coulomb" interaction is written as

\begin{equation}
V_C = \frac{1}{2}\int d^3x\, d^3y\, {\cal J}^{-1/2} \rho^A(\x)
 {\cal J}^{1/2} \hat K_{AB}(\x,\y;\A) {\cal J}^{1/2} \rho^B(\y) {\cal J}^{-1/2},
\label{eq:hc}
\end{equation}
where the Faddeev-Popov determinant is written as ${\cal J} \equiv {\rm det}(\bnabla\cdot \D)$ and $\D$ is the adjoint covariant derivative, 
 $\D^{AB} \equiv \delta^{AB} \bnabla  - g f^{ABC}\A^C$.
The color charge density is given by

\begin{equation}
\rho^A({\bf x}) =
 f^{ABC} {\bf A}^B({\bf x}) \cdot {\bf \Pi}^C({\bf x}) + \psi^{\dag}(\x)T^A\psi(\x).
\label{eq:rho}
\end{equation}
The kernel of the Coulomb interaction can be formally written as\cite{Christ:1980ku}

\begin{equation}
\hat K^{AB}({\bf x},{\bf y};\A) \equiv \langle{\bf x},A|
 \frac{ g }{ \bnabla\cdot {\bf D}}(-\bnabla^2)
 \frac{ g }{ \bnabla\cdot {\bf D}}|{\bf y},B\rangle.
\label{eq:K}
\end{equation}
Finally, $\A$ is the vector potential and ${\bm \Pi}$ is the conjugate momentum given by the negative of the transverse chromoelectric field.

The Coulomb interaction, along with the quark and gluon kinetic energies, gluon self-interactions, and the quark-transverse gluon interaction, $-g \int d^3x \psi^\dagger \bm{\alpha}\cdot \A \psi$, comprise a full field-theoretic version of QCD, with its accompanying nonperturbative features.

A quasigluon that is consistent with the constraints of QCD can be developed by making a mean field model of the gluonic vacuum. The ensuing Schwinger-Dyson equations can be truncated and solved to obtain estimates for the vacuum expectation of the kernel, $\hat K^{AB}$, and for the gluon dispersion relationship\cite{Szczepaniak:2001rg}. Here we choose to accept standard constituent quark model phenomenology and lattice results for the static quark interaction, and model the vacuum expectation of the Coulomb kernel as a confining potential:

\be
\langle \hat K^{AB}(\r,\A)\rangle \to
\delta^{AB}\left( -\frac{3}{4} \mathcal{C} + \frac{a_S}{r} - \frac{3}{4} \sigma r\right).
\label{eq:V}
\ee
Of course this reproduces the successes of the Cornell potential in nonrelativistic quark models. 
Higher terms in the $n$-body expansion of $\hat K$ can be incorporated in the formalism as required. 

The vacuum model also gives rise to a quasigluon that can be described by a field expansion parameterized with a dispersion relationship, $\omega = \omega(k)$.
Direct computation in the vacuum Ansatz yields an expression that is well-approximated by\cite{Szczepaniak:2001rg}

\be
\omega^2 = k^2 + m_g^2{\rm e}^{-k/b_g}
\label{eq:omega}
\ee
where the dynamical gluon mass is $m_g \approx 600$ MeV and the parameter $b_g \approx 6000$ MeV. We stress that the gluon remains transverse and properties, such as Yang's theorem, remain in place. Other vacuum Ans\"{a}tze are possible, for example a Gaussian wavefunctional (equivalent to the mean field approximation described) can be combined with the Faddeev-Popov operator, which gives rise to a dispersion relation that is well-described by the Gribov form, $\omega^2 = k^2 + m_g^4/k^2$\cite{Feuchter:2004mk}. 

The model can be validated by computing the excited gluonic potentials in the case of fixed color sources. Doing so reveals that the potential surfaces are not ordered according to lattice results. The discrepancy can be corrected by including tri-linear 
gluonic terms in $\hat K^{AB}$ in the computation\cite{Szczepaniak:2006nx}. These contributions are zero for the lowest surface,  which are dealt with exclusively in this work, thus trilinear couplings are neglected.

The resulting model can be thought of as a minimal extension of the constituent quark model with the addition of constituent gluon degrees of freedom and possible additional couplings (for example, the tri-linear gluon coupling or the gluon--Coulomb interaction).

\subsection{Hybrid States}

As is traditional, it is assumed that hybrid mesons are dominated by Fock states  with the lowest number of constituents. This approximation is unreasonable in perturbative QCD but is made plausible here by the relatively large quasigluon mass. As stated above, one of the goals of this study is to test this statement. 

It is convenient to construct the total gluon spin, $j_g$,  by 
coupling the gluon spin projection to the gluon angular momentum, $\ell_g$. Converting to the gluon helicity basis and assuming  that  $\ell_g = j_g$ reduces the product of two Wigner matrices to one and produces a factor of

\be
\chi^{(-)}_{\lambda,\mu} \equiv \langle 1 \lambda \ell_g 0| \ell_g \mu\rangle =
\begin{cases} 0, \ell_g = 0 \\ \frac{\lambda}{\sqrt{2}} \delta_{\lambda,\mu}, \ell_g \geq 1 \end{cases}.
\ee
This represents a transverse electric (TE) gluon in our model and forms the explicit realization of the axial constituent gluon. Alternatively, one may set $\ell_g = j_g\pm 1$ and obtain a transverse magnetic (TM) gluon with a Clebsch factor given by $\chi^{(+)}_{\lambda,\mu} = \delta_{\lambda,\mu}/\sqrt{2}$. Here we work exclusively with low lying TE hybrid mesons.

Combining with quark spins yields the final expression for a hybrid creation operator

\begin{align}
&|JM [LS \ell j_g \xi]\rangle = \frac{1}{2} T_{ij}^A\int \frac{d^3q}{(2\pi)^3}\, \frac{d^3k}{(2\pi)^3} \,
\Psi_{j_g;\ell m_\ell}({\bf k}, {\bf q})\, \sqrt{\frac{2 j_g +1}{4\pi}} \, D_{m_g\mu}^{j_g*}(\hat k) \, \chi^{(\xi)}_{\mu,\lambda} \nonumber \\
& \times
\langle \frac{1}{2} m \frac{1}{2} \bar{m} | S M_{S} \rangle \,
\langle \ell m_\ell, j_g m_g| L M_L\rangle \,
\langle S M_S, L M_{L} | J M \rangle \,
b_{{\bf q}-\frac{{\bf k}}{2},i,m}^\dagger \,
  d_{-{\bf q}-\frac{{\bf k}}{2},j, \bar{m}}^\dagger \,
a^\dagger_{{\bf k}, A, \lambda} |0\rangle.
\label{eq:PSI}
\end{align}
Quark and gluon particle operators are understood to create quasiparticles and $|0\rangle$ refers to the correlated vacuum discussed above.

By construction, the hybrid state is an eigenstate of parity and charge conjugation with eigenvalues given by

\be
P=\xi (-1)^{\ell + j_g+1}  \ \ \textrm{and} \  \ C= (-1)^{\ell+S+1}.
\ee


\subsection{Glueball States}

A reasonably large quasigluon mass encourages modelling charge-conjugation positive (negative) glueballs as two (three) quasigluon states. Again, it is preferable to work in the helicity basis, where much of the algebra simplifies. Combining two gluons into states with good parity and total angular momentum $J$ can be achieved with\cite{Szczepaniak:2003mr}

\be
|JM;\eta\rangle = \frac{1}{\sqrt{2}} \left( |JM;\lambda,\lambda'\rangle + \eta |JM; -\lambda,-\lambda'\rangle\right),
\ee
for which  $P|JM;\eta\rangle = \eta (-)^J \,|JM;\eta\rangle$ and $\eta = \pm 1$.

The helicity states are constructed as  

\be
|JM;\lambda,\lambda'\rangle = \frac{1}{\sqrt{2(N_c^2-1)}} \sqrt{\frac{2J+1}{4\pi}} \, \int \frac{d^3k}{(2\pi)^3} \, \psi(k) \, D^{J*}_{M,\lambda-\lambda'}(\phi,\theta,-\phi)\, \Pi \, a^\dagger(\k,\lambda,A) a^\dagger(-\k,\lambda,A) |0\rangle.
\ee
The number of colors is denoted $N_c$, $A$ is an adjoint color index, and $\Pi$ is a Jacob-Wick phase that will not be important to the following development. The wavefunction $\psi$ is determined by solving the Tamm-Dancoff equation that is obtained by evaluating the QCD Hamiltonian in the appropriate $J^P$ channel. This yields the leading order contribution involving the Coulomb interaction, Eq. \ref{eq:hc}. Higher order contributions from gluon exchange and the four-gluon interaction can be incorporated if desired.  The resulting spectrum is reported in Ref. \cite{Szczepaniak:2003mr}, where it is compared to lattice field theory computations. The spectra agree quite well where they overlap, with the largest deviation being about 200 MeV.

\section{Light Hybrid Spectrum}
\label{sect:3}

\subsection{Model Parameter Selection}
\label{sect:mm}

Because the primary goal is to compute hybrid mixing masses and angles, it is not necessary to obtain a precise hybrid spectrum. This is convenient because very little is known experimentally and because lattice field computations at physical pion masses and with coupled channel effects are not yet available. We therefore focus on spin-averaged hybrid multiplets in the following. In practice this means neglecting transverse gluon exchange contributions to the quark and gluon interactions. Thus the model parameters are the quark mass, the Coulomb coefficient, the string tension, and the constant shift: $m$, $a_S$, $\sigma$, $\mathcal{C}$. Recall that the gluonic parameters $m_g$ and $b_g$ have been fixed by the vacuum model. Including transverse gluons will introduce the coupling $g$ as well, which can be fixed by $a_S = g^2/(4\pi)$ or from other considerations to be discussed.

Model parameters will be fixed by fitting to 16 light isovector nonexotic  meson masses. As a check of stability we also fit to 30 isospin 0, 1/2, and 1 light mesons whose identities are reasonably well established. This is not necessarily a simple procedure since identifying ``non-canonical" properties in the light mesons is notoriously difficult. Famous examples include the $f_0(500)$ which has come and gone in the Review of Particle Physics (PDG) over the years. Similarly, the $a_0(980)$ has been identified as a $q\bar{q}$ state, a tetraquark, or a $K\bar{K}$ bound state by various authors. Lastly, the pion is a pseudo-Goldstone boson, hence simple quark models cannot be expected to reproduce its properties.

We do not presume to have a definitive description of light mesons and therefore will fit several model variations to obtain a sense of parameter stability in the subsequent work. These models are (i) spin-independent interaction (Eq. \ref{eq:V}) with a smeared hyperfine interaction, (ii) spin-independent interaction, (iii) spin-dependent interaction, (iv) variation on (iii), (v) spin-dependent interaction fit to 30 mesons. For the sake of comparison, results for models with relativistic quark kinetic energies are also given below, although these are not used in the subsequent analysis.

The hyperfine interaction used in model (i) is  given by

\be
V_{H} = \frac{32 \pi a_S}{9 m_q m_{\bar q}} \left(\frac{\varsigma}{\sqrt{\pi}}\right)^3\, \exp(-\varsigma^2 r^2)\,
\S_q\cdot \S_{\bar q},
\ee
with the smearing parameter set to $\varsigma =0.897$ GeV. For the other models, the spin-dependent interaction is defined by $V_{SD} =V_{H} + V_{LS} + V_{T}$ with:

\be
V_{H} = \frac{8 \alpha_H}{3} b_0^2 \frac{\textrm{e}^{-b_0 r}}{3m_q m_{\bar q} r}
\S_q\cdot \S_{\bar q},
\ee

\be
V_{LS}  = \left(\frac{4\alpha_H}{3 \rho^3} + \frac{\epsilon \sigma}{\rho}\right) \frac{\L\cdot \S}{m_q m_{\bar q}}  + \frac{1}{2}\left( \frac{4}{3}\frac{\alpha_H}{\rho^3} + \frac{(2\epsilon-1)\sigma}{\rho}\right) \, \left( \frac{\L\cdot \S_q}{m_q^2} + \frac{\L\cdot \S_{\bar q}}{m_{\bar q}^2} \right),
\ee
and

\be
V_T = \frac{4\alpha_H}{3 m_q m_{\bar q} \rho^3} \left( \S_q \cdot \hat r \, \S_{\bar q}\cdot \hat r - \S_q \cdot \S_{\bar q}\right).
\ee
The ultraviolet singularity in these expressions has been regulated by freezing $r$ at $r_0$ once $r<r_0$; this is denoted as $\rho$ in the equations. The parameter $\epsilon$ that appears in the spin-orbit tensor interaction represents a mixture of ``scalar" and ``vector" confinement models. Lattice computations find that $\epsilon \approx 0.25$\cite{KK}, which is used in model (v).

Results for the fits are presented in Table \ref{tab:models} and will be used to model hybrid mesons.

\begin{table}[h]
\caption{Model Parameters for the Isovector  Meson Spectrum.}
\begin{tabular}{l|cccccccc|cc}
\hline\hline
model & $m\ (m_s)$ (MeV) & $a_S$ & $\sigma$ (GeV$^2$) & $\mathcal{C}$ (MeV) & $\alpha_H$ & $b_0$ (GeV$^{-1}$) & $r_0$ (GeV$^{-1}$) & $\epsilon$ (GeV$^{-1}$) & rel error & avg deviation (MeV)\\
\hline
i.   [hyp] & 335 & 0.59 & 0.16 & -697 & -- & -- & -- & 0 & 9\% & 94 \\
ii.  [SI] & 300 & 1.52 & 0.071  & 110  & -- & -- & -- & 0 & 7\% & 66\\
iii. [SD] & 400  & 1.8  & 0.06  & 230  & 1.3  & 0.60 & 4.7 & 0 & 5\% & 54 \\
iv.  [SD] & 330  & 2.1  & 0.055  & 385  & 1.3  & 0.53 & 5.2 & 0 & 5\% & 54 \\
v.   [30] & 375 (525) & 1.3  & 0.059  & 112  & 0.77  & 0.84  & 5.2 & 0.25 & 7\% & 77 \\
\hline
Rel/SI & 200 & 0.59 & 0.14  & -246  & --  & -- & -- & 0 &  6\% & 59 \\
Rel/SD & 400 & 0.72 & 0.14  & -359 & 1.1 &  0.20 & 4.4 & 0 &  5\% & 55 \\
\hline\hline
\end{tabular}
\label{tab:models}
\end{table}

\subsection{Hybrid Mesons}
\label{sect:hm}

Spin-independent hybrid wavefunctions are obtained by considering the nonrelativistic limit of the interaction of Eq. \ref{eq:hc} (with Eq. \ref{eq:V}). The resulting spectrum can be categorized according to interpolating operators, as indicated in Table \ref{tab:JPC}\cite{JKM}. Here $\B$ is the chromomagnetic field, and $\psi$ and $\chi$ are heavy quark and antiquark fields, respectively. The remaining columns give the corresponding quantum numbers in the present model and the hybrid meson quantum numbers in the specified multiplet.

\begin{table}[h]
\caption{$J^{PC}$ Hybrid Multiplets.}
\begin{tabular}{c|c|cccc|l}
\hline\hline
multiplet & operator & $\xi$ & $j_g$ & $\ell$ & $L$ & $J^{PC}\ S=0\ (S=1)$ \\
\hline
$H_1$ & $\psi^\dagger \B \chi$ & -1 & 1 & 0 & 1 & $1^{--}$, $(0,1,2)^{-+}$ \\
$H_2$ & $\psi^\dagger \bnabla \times \B \chi$  & -1 & 1 & 1 & 1 & $1^{++}$, $(0,1,2)^{+-}$ \\
$H_3$ &$\psi^\dagger \bnabla \cdot \B \chi$ & -1 & 1 & 1 & 0 & $0^{++}$, $(1^{+-})$ \\
$H_4$ & $\psi^\dagger [\bnabla \B]_2 \chi$ & -1 & 1 & 1 & 2 & $2^{++}$, $(1,2,3)^{+-}$ \\
\hline\hline
\end{tabular}
\label{tab:JPC}
\end{table}

The bound state equation is obtained from the model QCD Hamiltonian by computing the expectation value, $\langle J'M'[L'S'\ell' j' \xi']| H | JM[LS \ell j \xi]\rangle$. As mentioned, we seek spin-independent multiplets and therefore consider the nonrelativistic limit of the currents in Eq. \ref{eq:hc}. This gives rise to instantaneous quark-antiquark and (anti)quark-gluon interactions that generate the bound state.

A novel method for solving the quantum mechanical three-body problem was applied to solve the resulting Schr\"{o}dinger equation. This consisted of writing the hybrid wavefunction as a sum over a product Ansatz of the form

\be
\Psi_{j_g;\ell m_\ell}({\bf k}, {\bf q}) = \chi_{j_g}(k)\, \varphi_{\ell}(q) Y_{\ell,m_\ell}(\hat q).
\label{eq:ansatz}
\ee
This makes explicit the angular momentum dependence in the $q$ coordinate, while the gluon angular momentum dependence is contained in the Wigner rotation matrix in Eq. \ref{eq:PSI}. 
In practice the basis used is nearly diagonal in the quantum numbers,  having only a coupling between the $H_1$ and $H_2$ multiplets induced by mixing between TM and TE hybrids (for spin-independent interactions). This will be small and is neglected.  It is thus possible to solve for $\varphi$ and $\chi$ separately and iterate the coupled equations to convergence. 

We sketch the idea here, ignoring all indices for simplicity. Write the Hamiltonian generically as $K_q + K_g +V$, where the first two terms are the quark and gluon kinetic energy operators obtained from the QCD Hamiltonian and the potential includes the sum over the three possible instantaneous interactions. Then vary $\langle \varphi\chi|H|\varphi\chi\rangle+ \lambda (\langle \varphi\chi|\varphi\chi\rangle-1)$ with respect to $\varphi$ and $\chi$. Eliminating the Lagrange multiplier yields

\begin{align}
K_q \varphi + \int \chi^* K_g \chi \cdot \varphi + \int \chi^* V \chi \cdot \varphi &= E \varphi  \nonumber \\
K_g \chi + \int \varphi^* K_q \varphi \cdot \chi + \int \varphi^* V \varphi \cdot \chi &= E \chi.
\label{eq:coupled}
\end{align}

Eq. \ref{eq:coupled} reduces to two one-dimensional equations. We solve this system by using a simple and accurate discretization method\cite{BS}, diagonalizing the Laplacian operator to deal with the (momentum space) gluonic kinetic energy\cite{BS}, and iterating. The latter step requires an initial guess for $\varphi$ and $\chi$, which is obtained variationally. In practice the method converges very quickly, and a precise  solution to the quantum bound state problem is obtained.

Calculations were done with the gluon dispersion relationship of Eq. \ref{eq:omega} with $m_g$ set to 600 MeV. Performing the same calculations with the Gribov form of the dispersion relationship and the same mass scale yielded very similar hybrid masses, with typical results being approximately 10 MeV heavier than given with Eq. \ref{eq:omega}.

\begin{figure}[ht]
\includegraphics[width=10cm,angle=0]{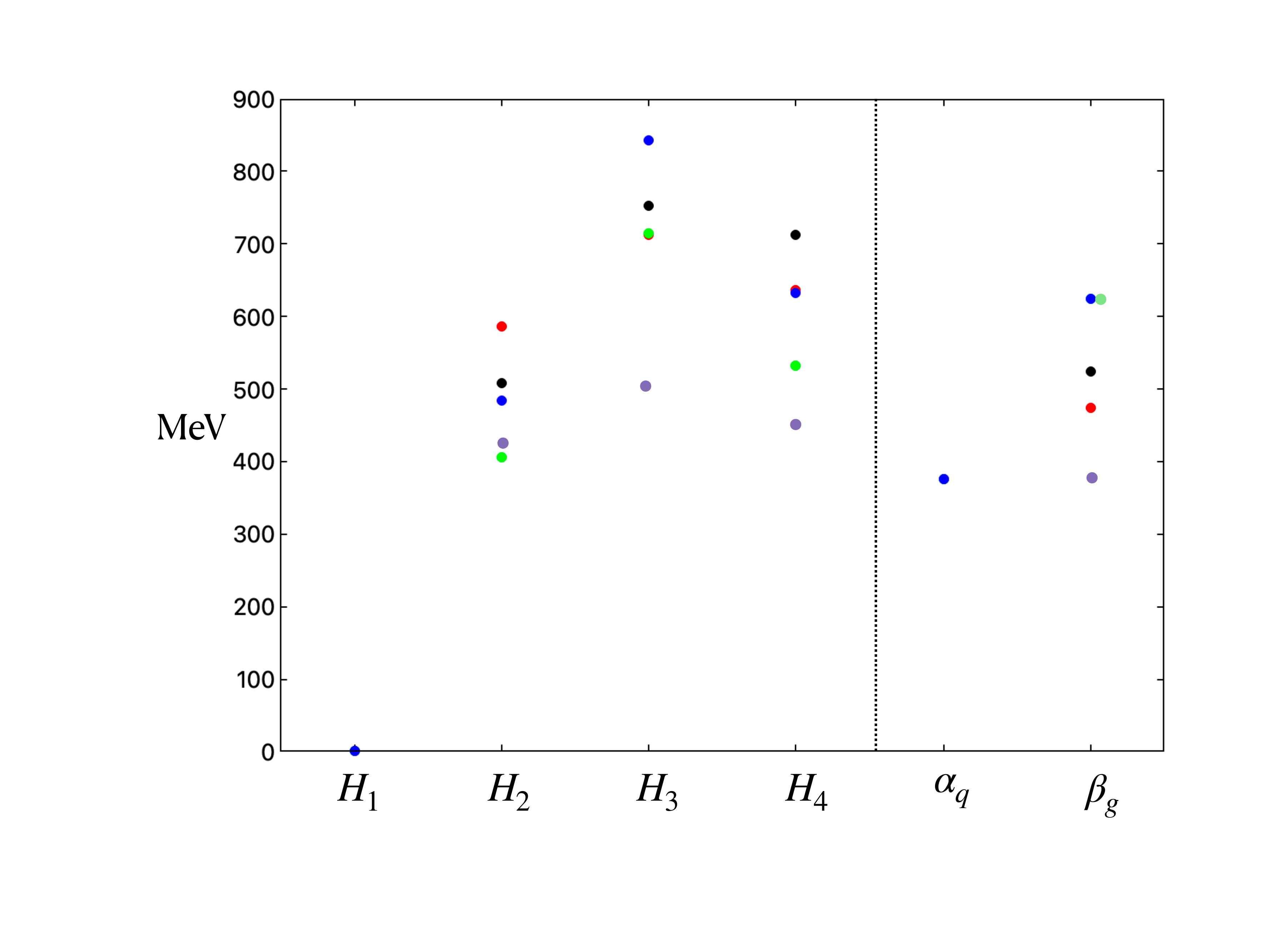}
\caption{Hybrid Multiplet Mass Splittings and $H_1$ Variational Parameters. Model i (red), ii (black), iii (green), iv (blue),v (purple).}
\label{fig:models}
\end{figure}

The resulting hybrid multiplet mass splittings with respect to $H_1$ are shown in Fig. \ref{fig:models}.
Detailed model validation is not feasible at present. A lattice computation of the light meson spectrum at a pion mass of 391 MeV exists, but it has not been able to distinguish enough hybrid states to determine the spin-averaged spectrum\cite{Dudek:2013yja}. This calculation does, however,  set the scale for the spin-averaged $H_1$ multiplet (taken to be the $S=0$ $1^{--}$ mass) to be  $2190\pm 20$ MeV.

On the experimental side, the situation is even more sparse and confused.  Past claims to  exotic $\pi_1$ states near 1400 and 1600 MeV\cite{Workman:2022ynf} have recently been challenged, with a consensus emerging that only one $\pi_1$ exists near 1600 MeV (results from two analyses are  $M= 1564 \pm 24\pm 86$ MeV, $\Gamma= 492 \pm 54 \pm 102$ MeV \cite{JPAC:2018zyd} and $M = 1623 \pm 47 \pm 50$ MeV, $\Gamma = 455 \pm 88 \pm 150$ MeV \cite{Kopf:2020yoa}).

Figure \ref{fig:models} displays a very large predicted splittings, $H_2 - H_1 \approx 500$ MeV.  This splitting can be estimated from the lattice calculation of Ref. \cite{Dudek:2013yja} using the $J^{PC} = 2^{+-}$ to $1^{--}$ mass difference (assuming that the $2^{+-}$ mass is approximately the $1^{+-}$ mass, which is supported by lattice calculations at the charmonium mass\cite{HadronSpectrum:2012gic,Cheung:2016bym}), yielding a value of approximately 250 MeV. Notice also that $H_3$ lies above $H_4$. This situation also occurs in a similar  model calculation for charmonium hybrids, reported in Ref. \cite{Farina:2020slb}, where it is seen to disagree with the lattice ordering, $H_4 > H_3$. Thus, although the model broadly agrees with lattice field theory calculations, where available, it is appears that additional effects, such as occur at higher order in the $1/m$ expansion, and further model tuning may be important to obtaining detailed agreement.

It is useful to find Gaussian estimates to the exact wavefunctions considered here so that an analytic evaluation of the mixing matrix elements (to be discussed in the next section) can be made. For this purpose we write $\chi_{j_g=1}(k) \propto k\exp(-k^2/\beta_g^2)$ and $\varphi_\ell(q) \propto q^\ell \exp(-q^2/\alpha_q^2)$. The parameters are estimated by optimizing the energy variationally and are shown to the right in Fig. \ref{fig:models} for the five models considered here.

\section{Hybrid Flavor Mixing}
\label{sect:4}

The topic of light meson flavor mixing is replete with experimental data, with much to be gleaned from a plethora of decay modes. In contrast, theoretical understanding of the issue is essentially nonexistent. The only certainty is that mixing occurs via nonperturbative gluodynamics, presumably dominated by coupling to intermediate glueballs.  Since so little is known of glueballs and their dynamics, theory is largely guesswork. (We remark that mixing in the $\eta-\eta'$ system is unique in that the axial anomaly makes a quantifiable nonperturbative contribution.)

%
%
%
%
%

In contrast to the rather grim situation with canonical mesons, hybrid mesons have their dominant quark configuration in a color octet, and can therefore mix perturbatively. The general situation in a given isospin multiplet involves amplitudes for mixing $u\bar u \leftrightarrow u\bar u$ and $u\bar u \leftrightarrow d\bar d$. These are expected to be nearly identical, hence both are labelled $A_{nn}$. Mixing $u\bar u \leftrightarrow s\bar s$ will be denoted $A_{ns}$, while $s\bar s \leftrightarrow s\bar s$ will be $A_{ss}$.

As we have mentioned, mixing to positive charge conjugation glueballs is first order in the strong coupling, and can therefore expected to be important. We label these amplitudes $\mathcal{A}^{(n)}_f$ where $f= n,s$ denotes the annihilated quark flavor and $(n)$ denotes a radially excited glueball of the relevant quantum numbers. We will show that this sum saturates quickly, so only the ground state glueball is considered in the following.

Diagonal elements of the QCD Hamiltonian will be written as $m$ for the $u\bar u$ and $d \bar d$ cases, $m+\Delta m$ for $s\bar s$, and $M_{gb}$ for the bare glueball mass. Thus, in the $u\bar u$, $d\bar d$, $s\bar s$ basis the matrix elements of the QCD Hamiltonian are

\be
H_{uds}= \begin{pmatrix} m+A_{nn} & A_{nn} & A_{ns} &  \mathcal{A}^{(0)}_n \\
A_{nn} & m+A_{nn} & A_{ns} &  \mathcal{A}^{(0)}_n \\
A_{ns} & A_{ss} & m + \Delta m + A_{ss} &  \mathcal{A}^{(0)}_s \\
\mathcal{A}^{(0)}_n  &  \mathcal{A}^{(0)}_n &  \mathcal{A}^{(0)}_s & M_{gb} 
\end{pmatrix}.
\label{eq:H}
\ee
Switching to the isospin basis $(u\bar u - d\bar d)/\sqrt{2}$, $(u\bar u + d\bar d)/\sqrt{2}$,  partially diagonalizes the mass matrix:

\be
H_{iso}= \begin{pmatrix} m & 0 & 0 & 0 \\ 
0 & m+2A_{nn} & \sqrt{2}A_{ns} & \sqrt{2}  \mathcal{A}^{(0)}_n \\
0 & \sqrt{2}A_{ns} & m + \Delta m + A_{ss} &  \mathcal{A}^{(0)}_s \\
0  &  \sqrt{2} \mathcal{A}^{(0)}_n &  \mathcal{A}^{(0)}_s & M_{gb} 
\end{pmatrix}.
\label{eq:H2}
\ee
A final diagonalization then gives the isovector, isoscalar, and glueball  masses and mixing angles.

\begin{figure}[ht]
\includegraphics[width=4cm,angle=0]{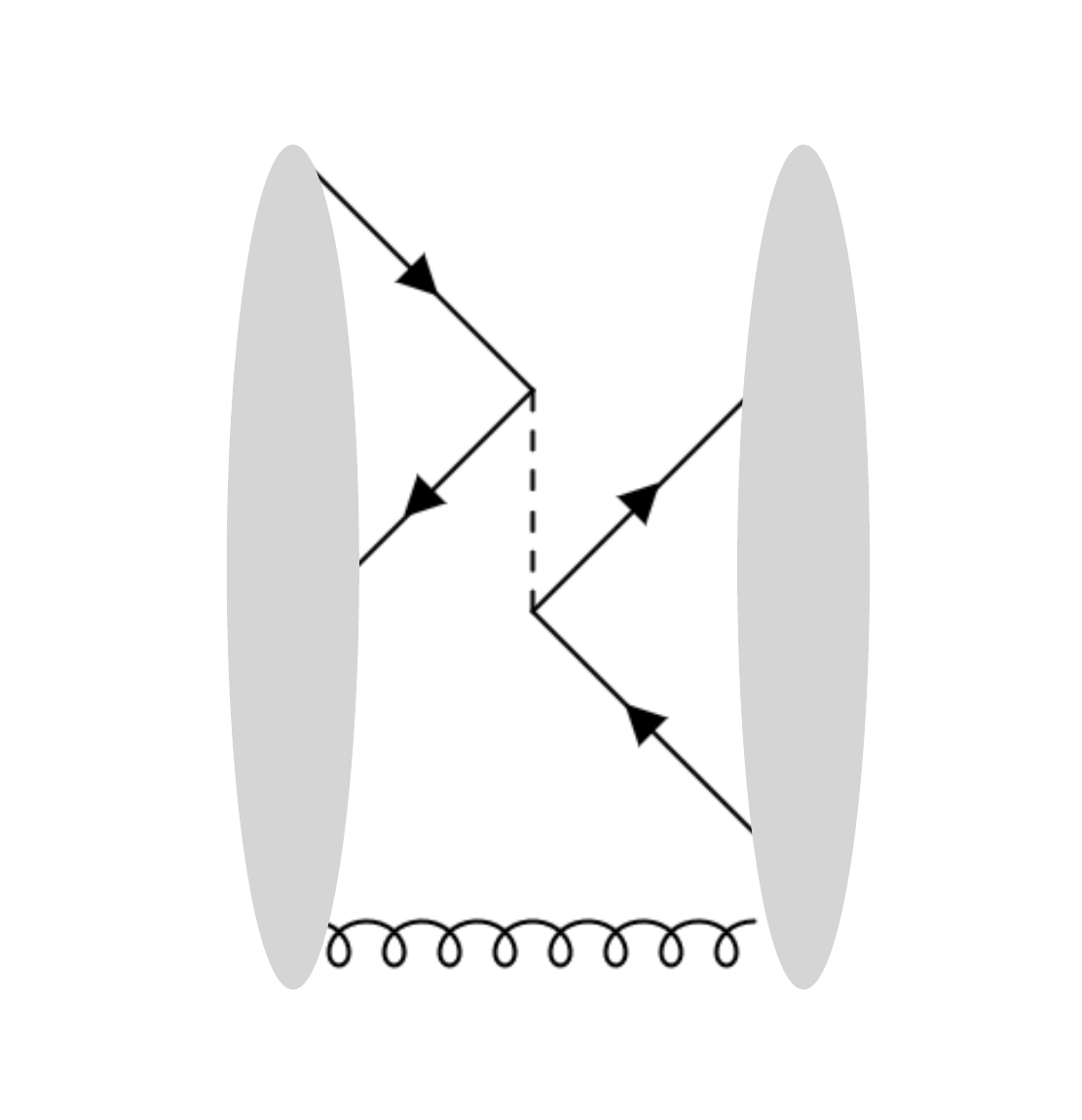}
\qquad
\includegraphics[width=4cm,angle=0]{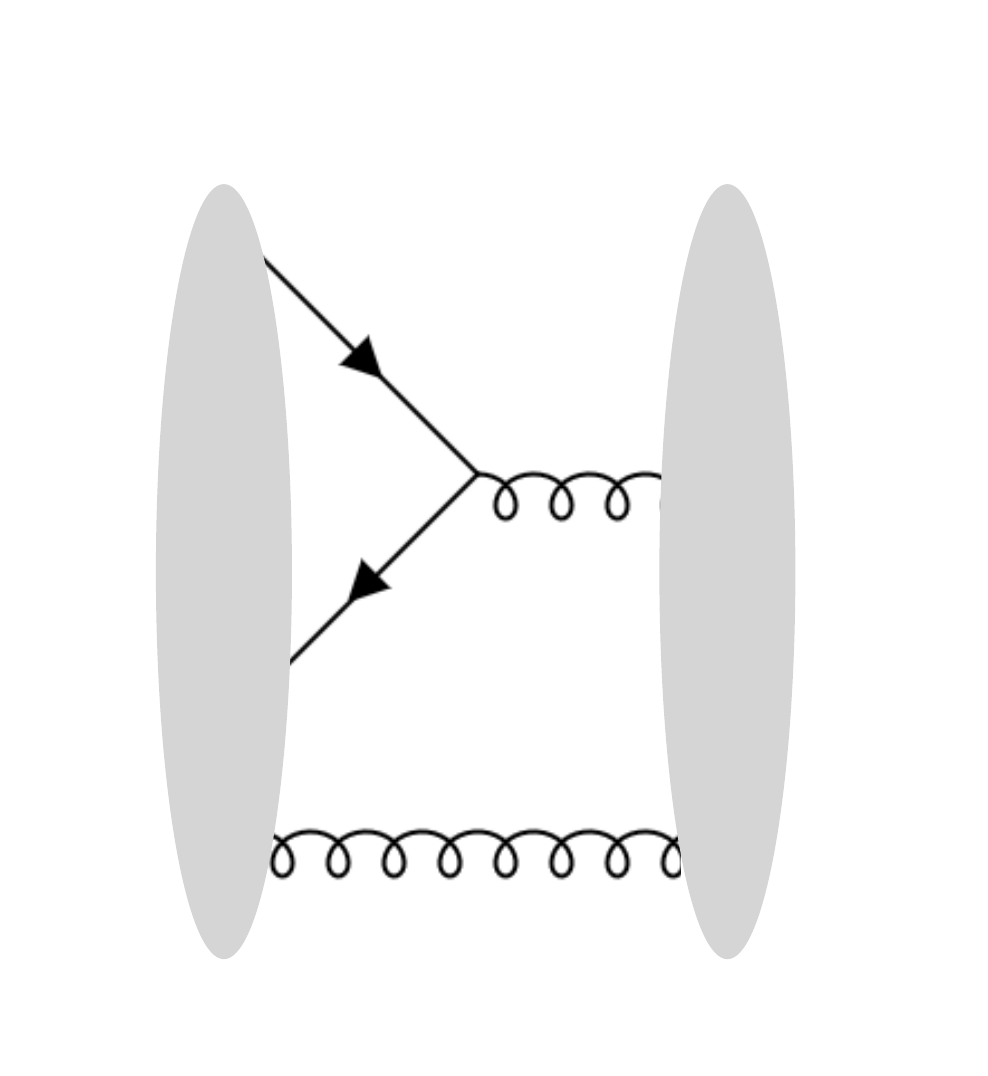}
\caption{(left) First order mixing diagram. (right) Hybrid-gluball mixing (crossed diagram not shown).}
\label{fig:m1}
\end{figure}

The leading (order $g^2$) sources of hybrid mixing are direct annihilation (Fig. \ref{fig:m1}, left), which occurs at first order in perturbation theory, second order mixing via glueball states (Fig. \ref{fig:m1}, right), or second order mixing to the meson-meson continuum. 
Flavor mixing via coupling to the meson-meson continuum has been enigmatic since the beginnings of the quark model. The issue, first stressed by Lipkin\cite{Zvi}, is that continuum mixing can vitiate the Okubo-Zweig-Iizuka (OZI) rule because a process such as $J/\psi \to D\bar D \to \omega$ is not suppressed. How the OZI rule arises in spite of this mechanism  has been explored by Geiger and Isgur, who argue that cancellations occur when all possible intermediate meson-meson channels are summed, giving rise to an emergent scale that is much smaller than $\Lambda_{QCD}$\cite{Geiger:1991ab,Geiger:1992va}. We shall assume that hybrid mixing via coupling to the meson-meson continuum is similarly suppressed in the case of hybrid states.

The leading order expression for the mixing amplitude, shown in Fig. \ref{fig:m1}, is given by

\be
A_{ff'} = \frac{1}{m_f m_{f'}} \int \frac{k^2dk}{(2\pi)^3}\, \frac{d^3q}{(2\pi)^3}\, \frac{d^3q'}{(2\pi)^3}\, \Psi_f(k,\q) \Psi_{f'}^*(k,{\q}') k^2 V(k)\, B_J = \frac{F_f F_{f'} }{8 m_f m_{f'}} \int \frac{k^2dk}{(2\pi)^3}\, |\chi_1(k)|^2 k^2 V(k)\, B_J.
\ee
The potential $V(k)$ is the Fourier transform of Eq. \ref{eq:V}. 
Wigner rotation matrices have been integrated and Clebsch-Gordan sums have been done to give the first form. The second follows from the product Ansatz of Eq. \ref{eq:ansatz} and introduces the ``octet decay constant" $F_f = \int \frac{d^3q}{(2\pi)^3} \varphi_{\ell=0}(\q)$, where implicit flavor-dependence is labelled by $f$. Evaluation of the discrete sums is considerably simplified because the quark vertex forces $S=S'=1$ and $\ell = \ell' =0$. Thus hybrid mixing at this order only exists in the spin-triplet portion of the $H_1$ multiplet with relative strengths given by the Clebsch factor 

\be
 B_J = \begin{cases}
0 & J=0\\
1 & J=1\\
3/5 & J=2
\end{cases}.
\ee
Finally, the integrals can be performed if Gaussian approximate wavefunctions are employed, giving

\be
A_{ff'} = \frac{\pi B_J}{2m_fm_{f'}} \left(\frac{\alpha_f \alpha_{f'}}{\pi}\right)^{3/2} [a_S+\frac{b}{\beta_g^2}],
\ee
where $\beta_g$ is the gluonic scale  introduced after Eq. \ref{eq:coupled}.

Second order mixing via intermediate glueballs can be computed with the amplitude of Fig. \ref{fig:m1} (right). This diagram also features quark-antiquark annihilation in the ${}^3S_1$ channel, and therefore mixing is predicted to be zero for hybrids in the $H_2$, $H_3$, and $H_4$ multiplets, as well as the light vector, $H_1(1^{--})$ multiplet. This striking observation can only be reasonably evaded by mixing to negative charge conjugation (three-quasigluon) glueballs, which is expected to be small, or by coupling to the meson-meson continuum, which remains an enigmatic feature of hadronic physics.

The expression for the amplitude coupling a hybrid to the $n$th radial glueball excitation is 

\be
\mathcal{A}^{(n)}_{f} = -\frac{i [g F_f]}{4} \, \int \frac{k^2 dk}{(2\pi)^3}\, \frac{\psi_n^*(k)\chi(k)}{\sqrt{\omega(k)}} \, C_J,
\label{eq:Agb}
\ee
where 
\be
C_J = \begin{cases} 
              4, &  J=0 \\
              0, &   J=1\\
              4/\sqrt{10}, & J=2
       \end{cases}.
\ee

Note that the octet decay constant, $F_f$, appears in the glueball amplitude, this time combined with the strong coupling constant. Evidently, obtaining an accurate estimate of the decay constant is important. This can be problematic because it is known that the nonrelativistic approximation over-estimates the value of (traditional) meson decay constants. Incorporating relativistic effects helps, but it appears that further softening is required. This softening can occur, for example, via the effect of the running coupling on the wavefunction at the origin\cite{Lakhina:2006vg}. These issues are exacerbated in the case of glueball mixing (Eq. \ref{eq:Agb}) because the strong coupling constant is explicit. One can set the value of the strong coupling from the model via $g = \sqrt{4\pi a_S}$. This has the appeal of consistency, but might not be optimal because $a_S$ is a model parameter that is determined by bulk light hadron properties that are dominated by distances near 1 fm. Alternatively, a decay constant is a short range phenomenon--roughly speaking we wish to evaluate $[g F_f] \sim \sqrt{\alpha_S(r=0)} \varphi(r=0)$. Of course this implies a nontrivial infrared fixed point for the running coupling, which will be assumed here. A better way to proceed is to write $[g F_f] \sim \alpha_S(Q_*) \varphi(r=0)$ where $Q_*$ is a scale that is tuned to the physics. An alternative, that is adopted here, is to implicitly fix the scale by averaging; thus we set

\be
[g F_f] = \int \frac{d^3q}{(2\pi)^3} \, \sqrt{4\pi \alpha_V(q)} \, \phi_{\ell=0}(q)
\label{eq:run}
\ee
where the running coupling is parameterized as

\be
\alpha_V(q) = \frac{4\pi}{b_0 \log(\frac{q^2+M^2}{\Lambda_V^2})}.
\ee
with $b_0 = 11-2 n_f/3 = 9$ for our case. This is a reasonably common model that has been advocated for renormalizing exclusive processes\cite{Brodsky:1997dh} and has been used in modelling heavy meson properties\cite{Lakhina:2006vg}. Parameters chosen in these studies were $M= 870$ MeV, $\Lambda_V = 160$ MeV or $M= 1000$ MeV, $\Lambda_V = 250$ MeV, respectively. Results for both model choices ($\sqrt{4\pi a_S}$ and $\sqrt{4 \pi \alpha_V}$) will be presented in the following section.


Lastly, we address the issue of the convergence of the hybrid mixing amplitude in the sum over glueball excitations.
The perturbative glueball mixing amplitude is given by

\be
A^{gb}_{ff'} = \sum_n \frac{{\mathcal{A}_{f}^{(n)}}^*\mathcal{A}^{(n)}_{f'}}{M_{hyb} - M^{(n)}_{gb}}.
\label{eq:A2}
\ee
This sum is expected to converge quickly because the integral in Eq. \ref{eq:Agb} rapidly decreases with radial quantum number. In fact, the integral would be zero (for $n>0$) in a simple harmonic approximation to the wavefunctions if the glueball and hybrid scales were the same. An explicit calculation in the $J^{PC}=0^{-+}$ case shows that $\mathcal{A}_n^{(1)} \approx 0.37 \mathcal{A}_n^{(0)}$. The corresponding term in $A^{gb}_{nn}$ is further suppressed by the larger radial glueball mass, giving a final contribution that is only 1\% of the leading term in the sum. Because of this we only considered hybrid coupling to the lowest mass glueball in a given $J^P$ channel in Eq. \ref{eq:H2}.

\section{Comparison to Lattice Computations}
\label{sect:5}

\begin{figure}[ht]
\includegraphics[width=12cm,angle=0]{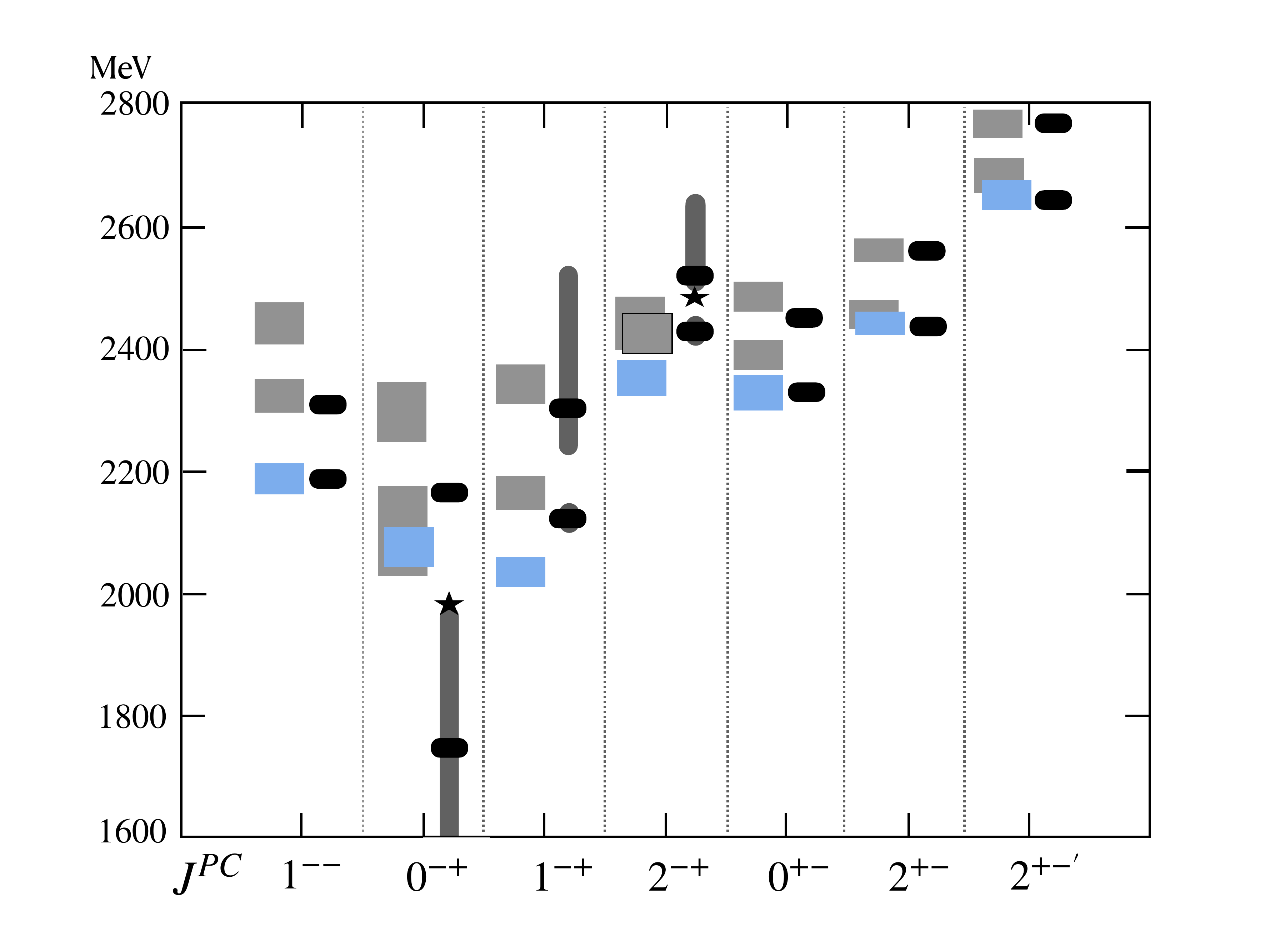}
\caption{Isovector and isoscalar hybrid masses. Lattice results are blue (isovector) and gray boxes (isoscalars). The box heights indicate statistical uncertainty. Model results for isoscalar masses are shown in black with model variation as vertical bars. Stars indicate model results with the running octet decay constant model of Eq. \ref{eq:run}.}
\label{fig:H}
\end{figure}

We proceed by diagonalizing the matrix of Eq. \ref{eq:H2} for the three nontrivial cases, $J^{PC} = 0^{-+}$, $1^{-+}$, and $2^{-+}$ (recall that all other hybrid mesons are predicted to have negligible mixing).  The resulting masses are shown in Fig. \ref{fig:H}, along with lattice masses computed in Ref. \cite{Dudek:2013yja}. The latter are computed at a pion mass of 391 MeV and therefore may experience some shifts in going to physical quark masses. In view of this, the model results have been normalized by setting $m$ to the isovector mass (shown in blue) for each $J^{PC}$ multiplet. 
Model isoscalar masses are also shown as black oblongs in the figure (states dominated by glueball components lie substantially higher and are not shown). Computing masses over the models of Table \ref{tab:models} gives an indication of parameter dependence. This dependence is indicated in the figure by vertical grey bars.

As discussed above, no mixing is expected outside the  $H_1$ multiplet. Evidently this prediction agrees very well with the lattice results for the $J^{PC}=0^{+-}$, $2^{+-}$, and $2^{+-'}$ multiplets. Countering this is the $1^{--}$ multiplet, where the lowest isoscalar is computed to be approximately 140 MeV above the isoscalar, rather than degenerate with it as predicted here. This  curious situation is difficult to reconcile with the current model. The dominant mixing effect would be via negative parity glueballs, which requires a gluon emission followed by a quark spin flip and then  quark pair annihilation. This process will be suppressed by a relatively large three-quasigluon glueball mass and the spin flip.

Turning attention to the $2^{-+}$ multiplet, reasonable agreement is seen for the lowest isoscalar mass, with some overlap for the higher isoscalar--especially for the preferred running coupling octet decay constant of Eq. \ref{eq:run}. In this case we suspect that the higher lattice mass is anomalously low since it is implausible for isoscalars to shift mass with respect to the isovector while remaining degenerate. In contrast, the $1^{-+}$ multiplet splits as expected and in reasonable agreement with the model calculation. 

Finally, the $0^{-+}$ multiplet is very unusual--the lattice results imply very small mixing, in contrast the $\sqrt{a_S}$ model predicts that the light isoscalar lies from 50 to several hundred MeV below the isovector. This is because there is no direct mixing, $A_{nn}(J^{PC}=0^{-+}) = 0$, and mixing with the pseudoscalar  glueball drives the light isoscalar mass down. The upward shift of the glueball is comparable to the downward shift of the light isoscalar, hence if the larger isoscalar shift proves correct, then unquenched lattice calculations of the pseudoscalar glueball mass should find it shifted several hundred MeV above the unquenched value of approximately 2600 MeV\cite{Chen:2005mg}. Of course the lattice results of Fig. \ref{fig:H} argue against this, and imply that the splittings are at the small end of the predicted range. Alternatively, the preferred running octet decay constant model gives a much smaller mixing that is reasonably close to lattice results--although still indicating a novel light isoscalar.

\begin{table}[ht]
\begin{tabular}{ll|ccc}
\hline\hline
$J^{PC}$ & nominal state & $u\bar u g$ & $s \bar s g$ & $gg$ \\
\hline
$1^{--}$ & light & $\approx 100$  & $\approx 0$ & $\approx 0$ \\
         & heavy & $\approx 0$ & $\approx 100$ & $\approx 0$ \\
         & glueball & $\approx 0$ & $\approx 0$  & $\approx 100$  \\
$0^{-+}$ & light &  62 [87] & 17 [6]  & 21 [7]\\
         & heavy &  24 [8] & 75 [91] & 0.5 [1] \\
         & glueball & 13 [5] & 8 [3] & 78 [92]\\
$1^{-+}$ & light & 37 & 63 & 0 \\
         & heavy & 63 & 37 & 0 \\
         & glueball & 0 & 0 & 100 \\
$2^{-+}$ & light & 54 [59]& 46 [41]& 0.1 [0]\\
         & heavy & 32 [38] & 67 [61]& 1 [1]\\
         & glueball & 2 [0.4]& 1 [0.6]& 97 [99]\\
\hline\hline
\end{tabular}
\caption{Model Hybrid $H_1$ Multiplet Fock Space Components (\%). Results for 
the running octet decay constant model of Eq. \ref{eq:run} are 
shown in square brackets.}
\label{tab:comp}
\end{table}

\begin{figure}[ht]
\includegraphics[width=12cm,angle=0]{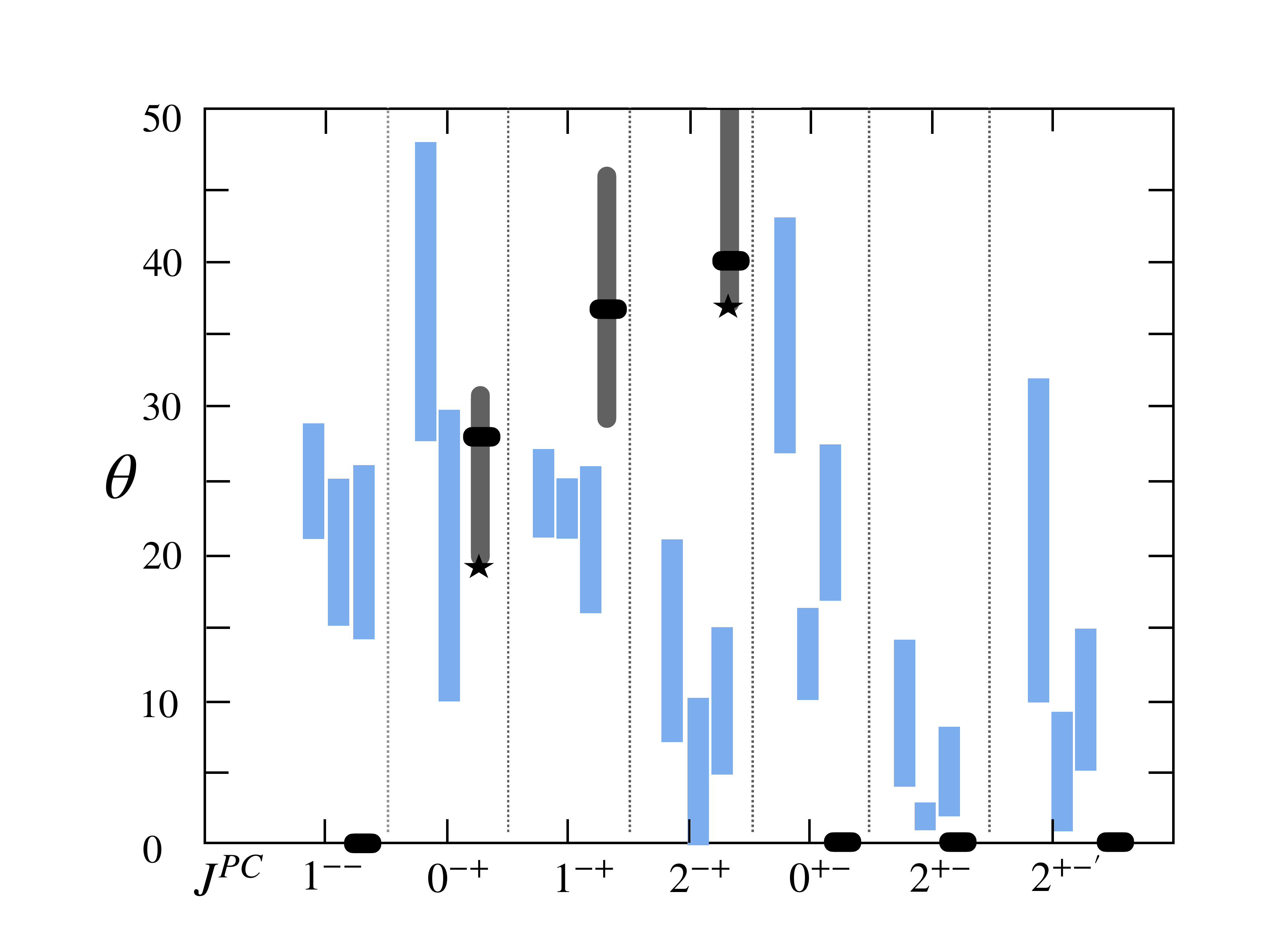}
\caption{Lattice and model mixing angles. Lattice mixing angles for three volumes($16^3$, $20^3$, $24^3$) with statistical errors indicated by the box height (blue). Model results (black) with model variation shown as vertical bars. 
Stars indicate model results with the running octet decay constant model of Eq. \ref{eq:run}.}
\label{fig:A}
\end{figure}

The authors of Ref. \cite{Dudek:2013yja} also report mixing angles between the two lightest isoscalar mesons. These were computed under the assumption that the states do not mix with nearby glueballs.  Of course this assumption is not made here; however, the erstwhile mixing angle can still be computed. This will lead to an ambiguous result if mixing to glueballs is substantial. As shown in Tab. \ref{tab:comp}, we find that this is not the case for all multiplets except $J^{PC}=0^{-+}$ (in the $\sqrt{a_S}$ model). Lattice mixing angles for three volumes are shown in blue in Fig. \ref{fig:A}; model results are displayed as black oblongs. As with the meson masses, the results are in broad agreement with lattice (in view of the large errors), with the largest discrepancy being in the vector multiplet again.

\section{Hybrid-Canonical Meson Mixing}
\label{sect:6}

The encouraging results of the previous section motivate the consideration of hybrid meson mixing with canonical mesons.  Work of this sort dates back to the beginnings of the bag model and QCD, see for example\cite{Barnes:1977hg}. For a computation based the Born-Oppenheimer approach, see Ref.~\cite{Gerasimov:1998tm}.

 Following the philosophy advocated here, this process will be mediated by gluon production from quark and anti-quark lines. Thus we seek $\mathcal{H} \equiv \langle q\bar q g|ig \int \psi^\dagger \bm{\alpha}\cdot\A\psi| q{\bar q}\rangle$. Configuration mixing of this sort is of most interest for vector states since it has implications on the coupling of vector hybrids to $e^+e^-$.

Taking the nonrelativistic limit of the vertex,  performing integrals over the angles of $\k$, and doing the Clebsch sums gives a result involving an integral of the hybrid wavefunction convoluted with the vector quarkonium wavefunction:

\be
\mathcal{H} = -i\frac{g}{m}\frac{2\sqrt{4\pi}}{3} \int \frac{d^3q}{(2\pi)^3} \, \frac{k^2dk}{(2\pi)^3}\, \frac{k}{\sqrt{\omega(k)}} \Psi^*(k,\q) \psi(\q+\k/2).
\ee

The hybrid wavefunction is obtained with the method of Sect. \ref{sect:hm} for $u\bar u g$, $c \bar c g$, and $b\bar b g$ $H_1$ vector hybrids. Wavefunctions corresponding to $\rho$, $J/\psi$, and $\Upsilon$ mesons were obtained as outlined in Sect. \ref{sect:mm}. 
The numerical results are

\be
\mathcal{H} = -i g\begin{cases} 84 \ \textrm{MeV}^2/m_q, & \rho \\
                               190 \ \textrm{MeV}^2/{m_c}, &  J/\psi \\
                               225 \ \textrm{MeV}^2/{m_b}, &  \Upsilon
                   \end{cases} \approx 
                  -i \begin{cases} 210 \ \textrm{MeV},  & \rho\\
                                60\ \ \textrm{MeV}, & J/\psi \\
                                20\ \ \textrm{MeV}, & \Upsilon
                   \end{cases}.
\label{eq:hqq}
\ee

Model validation is not simple because phenomenological mixing information is not available (since hybrid mesons of any type are not firmly established). Comparison to lattice field theory is difficult because the formalism automatically produces eigenstates over the $q\bar q$ and $q\bar q g$ Fock state sectors. There is, however, one lattice computation (that I am aware of) that uses the nonrelativistic QCD (NRQCD) formalism. This permits defining bare Fock states and measuring their overlap, in this case due to the operator $g \bm{\sigma}\cdot \B/(2m)$\cite{Burch:2003zf}. The computation is not easy (the authors note, ``Our charmonium results are plagued with systematic errors which are not easily quantified"), involving a poorly determined renormalization constant, and difficulties in scale setting. Nevertheless, the authors estimate a hybrid component of approximately 2.3\% in the $J/\psi$ and 0.4\% in the $\Upsilon$. Approximating these as $|\mathcal{H}/(M_{hyb}-M_V)|^2$ and using the measured charmonium vector mass splitting of approximately 1150 MeV\cite{HadronSpectrum:2012gic}, gives

\begin{eqnarray}
\mathcal{H}_{\scriptstyle NRQCD} &\approx& 170 \ \textrm{MeV}\ (J/\psi) \nonumber \\
 &\approx& 70\ \textrm{MeV}\ \ (\Upsilon). \nonumber
\end{eqnarray}
The results of Eq. \ref{eq:hqq} are approximately a factor of three below these. Nevertheless, both computations contain large unquantified uncertainties, and it is encouraging that they are comparable in size and that the ratio of results does not follow the naively expected inverse quark mass relationship for either calculation. Over all, we take these results as evidence in favor of the utility of the constituent gluon/Coulomb gauge model presented here.  Application of the formalism to the isovector vector mesons will be presented in the following section.

\section{Hybrid Phenomenology}
\label{sect:7}

We examine the impact of the results presented here on interpreting the light meson spectrum, with a focus on vector states since these can be made in $e^+e^-$ machines.  It is unfortunate that experimental knowledge of the excited rho spectrum is spotty. For example, the $\rho(1450)$ and $\rho(1700)$ are both seen in $4\pi$ or $a_1\pi$ decay modes, which is a signal for hybrid structure. Clarifying the situation with further experimental and lattice field effort is clearly of interest. Similarly, the $\rho(1900)$ region has conflicting signals and complications from $N\bar N$ threshold\cite{Workman:2022ynf}. Finally, the BaBar collaboration has measured $e^+e^- \to 2(\pi^+\pi^-)\pi^0$, which reveals interesting (although low statistics) features near 2100 MeV\cite{BaBar:2007qju}.

\begin{table}[h]
\caption{Model Assignments and Experimental Isovector Vector States (MeV).}
\begin{tabular}{llll|ll|ll}
\hline\hline
state & Ref. & mass & width & model (iv) & mass & model (v)  & mass \\
\hline
$\rho(770)$ & PDG  & $775.2\pm 0.2$ & $147.4 \pm 0.8$ &  $1{}^3S_1$ & 720  &   $1{}^3S_1$  & 810 \\
$\rho(1450)$& PDG  & $1465 \pm 25$  & $400 \pm 60$   &  $2{}^3S_1$ & 1440   &   $2{}^3S_1$ & 1405 \\
$\rho(1570)$& PDG  & $1570 \pm 70$  & $144 \pm 90$   &  $1{}^3D_1$ & 1510   &  $1{}^3D_1$ &  1497 \\
$\rho(1700)$& PDG  & $1720 \pm 20$  & $250 \pm 100$  &  $H_1(1^{--})$ & 1760 &   $3{}^3S_1$  & 1770 \\
         &      &                &                &  &  & $2{}^3D_1$   & 1835 \\
         &      &                &                &  $3{}^3S_1$ & 1850  &  &   \\
$\rho(1900)$ & \cite{BaBar:2007ceh} & $1900 \pm 30$   & $50 \pm 30$  &  $2{}^3D_1$ & 1910  &  $4{}^3S_1$  &  2080  \\
$\rho(2150)$& \cite{BESIII:2020xmw} & $2034\pm 16$   & $234\pm 39$ &  $4{}^3S_1$ & 2170 &  $H_1(1^{--})$  &  2100  \\
         &      &                &                &  $3{}^3D_1$ & 2220 &  $3{}^3D_1$  & 2130 \\
\hline\hline
\end{tabular}
\label{tab:rhos}
\end{table}

\begin{figure}[ht]
\includegraphics[width=10cm,angle=0]{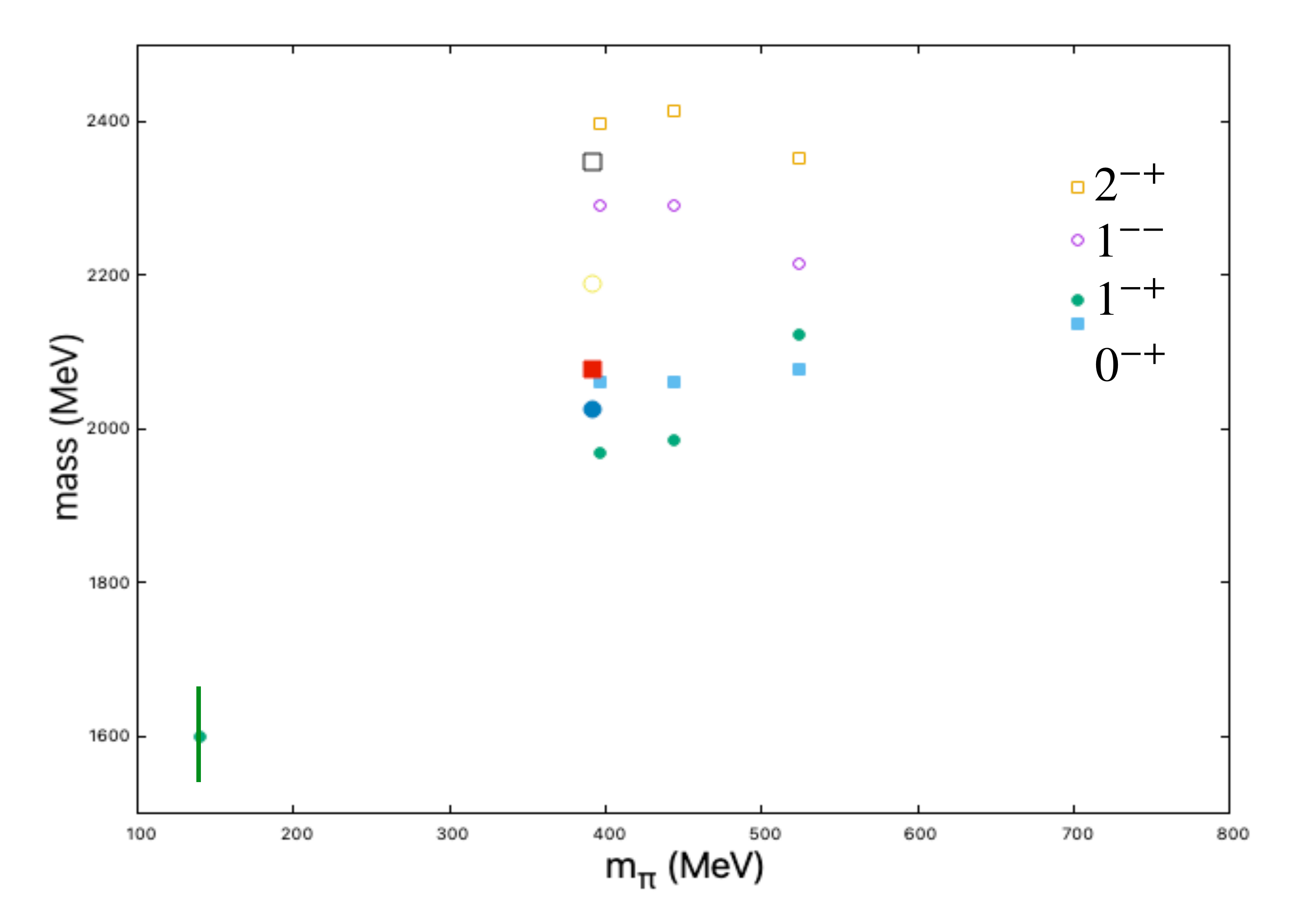}
\caption{Lattice $H_1$ Masses\cite{Dudek:2011bn} (large symbols Ref.~\cite{Dudek:2013yja}) vs. Pion Mass. The $\pi_1$ is indicated at lower left.} 
\label{fig:extrap}
\end{figure}

A summary of possible quark model identifications for the rho spectrum is shown in Table \ref{tab:rhos}. Models (iv) and (v) masses are in rough agreement, but notice that deviations of tens of MeV low in the spectrum become 90 MeV by the $4S$ state. This, seemingly minor, difference can lead to substantial changes in interpretation. For example, there is no vector state near 1700 MeV in model (iv), raising the possibility that the $\rho(1700)$ may be a hybrid state. The large symbols in Figure \ref{fig:extrap} are from a lattice computation at $m_\pi = 391$ MeV\cite{Dudek:2013yja}. The $1^{-+}$ state is measured at 2026 MeV, requiring a shift of 430 MeV to bring it to agreement with a presumed $\pi_1(1600)$. This then implies that the $H_1(^{--})$ should have a mass of 1760 MeV. In view of this is it tempting to make the particle identifications shown in column 5 of Table \ref{tab:rhos}.

The smaller points in Fig. \ref{fig:extrap} (obtained from Ref. \cite{Dudek:2011bn}) show $H_1$ hybrid masses computed at $m_\pi = 702$, 524, 444, and 396 MeV. These data permit a rough extrapolation to the physical light quark mass (evidently the data are not in the chiral regime, so a simple extrapolation in $m_\pi$ is used here).  We estimate a $H_1(1^{-+})$ mass of 1750 MeV, which implies an additional shift of 150 MeV. Performing the same procedure for the $H_1(1^{--})$ gives an estimated physical mass of 2100 MeV for the vector hybrid. A natural candidate for this state is the $\rho(2150)$. The $3S$ mass in model (v) matches the $\rho(1700)$ well, and we arrive at the alternative scenario spelled out in the last two columns of Table \ref{tab:rhos}.

There is a clear moral to this story: particle identification relies on correctly interpreting mass differences of order 100 MeV. Both lattice and model variations can easily generate deviations of this magnitude, thus it is important to track model sensitivity in making assignments.

In spite of the lack of moral clarity, it is likely that model (v) is more robust than model (iv) because it is fit to a much larger array of meson masses. The model (iv) scenario also has a missing $3S$ state, which is more unlikely than the missing $2D$ state of model (v).
It also seems clear that the lattice $1^{-+}$ mass has a larger quark mass-dependence than the $1^{--}$ mass. Lastly, as mentioned above,  intriguing structure is seen in 
 $e^+e^- \to 2(\pi^+\pi^-)\pi^0$ near 2100 MeV\cite{BaBar:2007qju}. In sum, we feel that the scenario of model (v) should be taken seriously.

If the $\pi_1(1600)$ is confirmed as a hybrid meson and a $\rho(2100)$ is found that matches expectations for hybrid production and decay, it is natural to inquire into the accompanying flavor multiplets that must exist. Taken together, the model and lattice results of Sect. \ref{sect:5} imply that an isoscalar $\eta_1$ should exist at 1750-1780 MeV, while the ``$s\bar s$" isoscalar should have a mass of approximately 1900 MeV. Interestingly, the BESIII collaboration recently announced the discovery\cite{BESIII:2022riz} of an exotic isoscalar meson with $J^{PC}=1^{-+}$ quantum numbers, a Breit-Wigner mass of $1855 \pm 9 \pm 4$ MeV, and a width of $188\pm 18 \pm 5$ MeV. Given the discussion it is natural to identify this state with the  $s\bar s$ isoscalar partner of the $\pi_1(1600)$, and suggests that searching for $\eta_1(1760)$ is in order.

\begin{figure}[ht]
\includegraphics[width=10cm,angle=0]{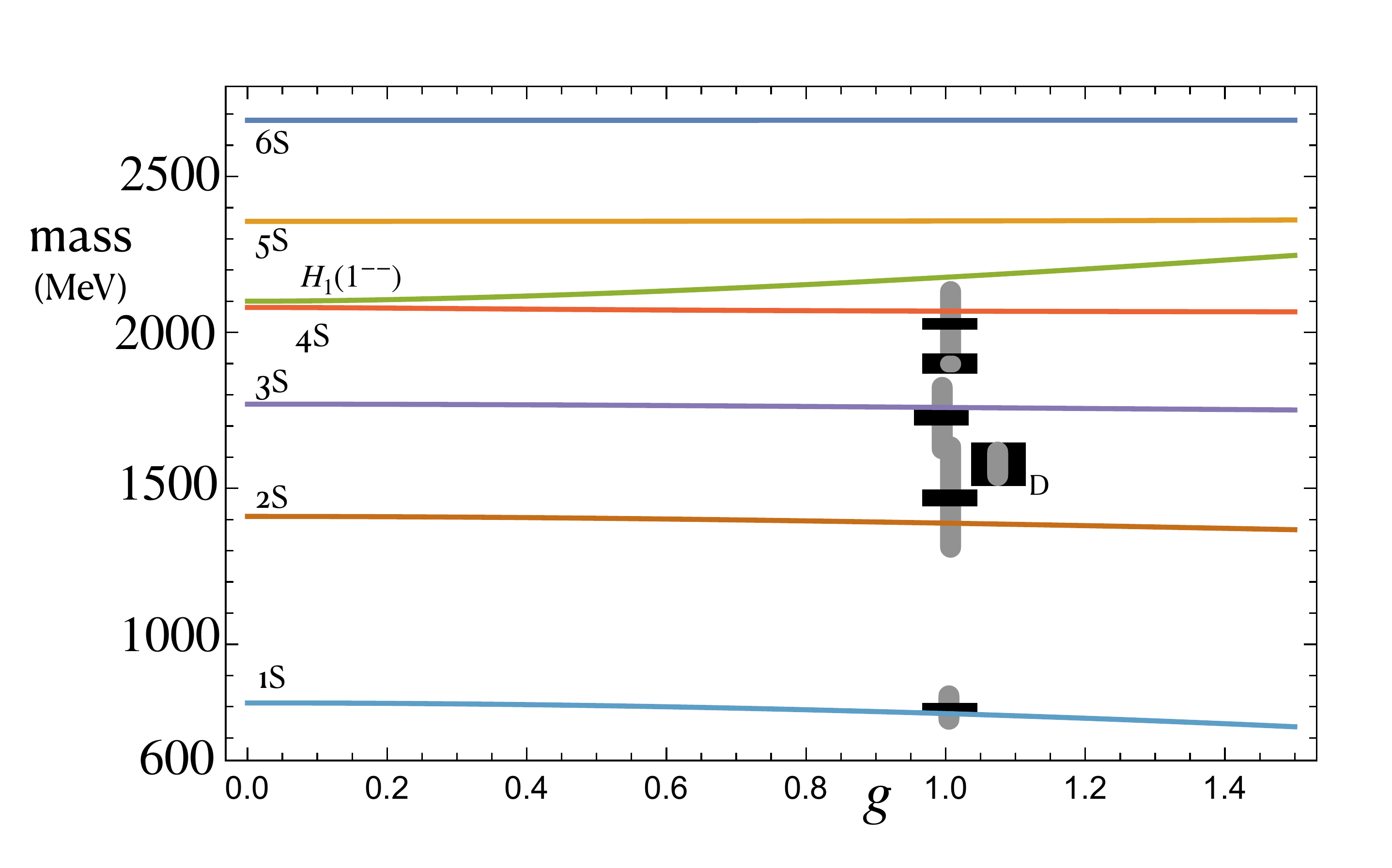}
\caption{Vector isovector masses with hybrid mixing as a function of the strong coupling. Boxes indicate experimental masses and their uncertainties. Grey bars indicate state widths.}
\label{fig:rhos}
\end{figure}

As discussed in Sect. \ref{sect:6}, the $H_1(1^{--})$ hybrid is expected to mix with nearby canonical vectors. This mixing is explored here to set expectations for the isovector vector hybrid, whose lattice mass we have suggested is near 2100 MeV. Using typical values for the strong coupling and light quark masses gave a mixing matrix element of $\mathcal{H}_{1S} \approx 210$ MeV for the $\rho(770)$ (see Eq. \ref{eq:hqq}). Repeating the calculation for radially excited rho mesons gives

\begin{eqnarray}
&&
\mathcal{H}_{1S} \approx 210\ \textrm{MeV} \qquad 
\mathcal{H}_{2S} \approx -130\ \textrm{MeV} \qquad 
\mathcal{H}_{3S} \approx 68\ \textrm{MeV} \nonumber \\
&&
\mathcal{H}_{4S} \approx -35\ \textrm{MeV} \qquad 
\mathcal{H}_{5S} \approx 16\ \textrm{MeV} \qquad 
\mathcal{H}_{6S} \approx -8\ \textrm{MeV}. \nonumber
\end{eqnarray}
The effect of this mixing on the spectrum is obtained by diagonalizing a matrix with the diagonals set by the quark model masses of model (v) (see Table \ref{tab:rhos}). (Strictly speaking bare quantities should appear in the mixing matrix. But the lattice hybrid mass (should) already account for mixing, while the quark model should absorb the effects of this mixing (where possible) in its parameters. However, as will be demonstrated next, mixing is small and hence this procedure serves as a useful illustration of the expected size of the effect).
The hybrid entry is set to 2100 MeV, while the off-diagonal entries are set to $\mathcal{H}_{nS}$ in the hybrid:$(nS)$ entry. The resulting spectrum is displayed versus the strong coupling in Fig. \ref{fig:rhos}. It appears that mixing effects are small, with the chief outcome being that the $\rho(770)$ and $H_1(1^{--})$ repel as the coupling is increased. Experimental masses and widths are displayed as boxes and vertical bars in the figure. The agreement between expectations and experiment is reasonable. Note especially that the $4S-H_1(1^{--})$ splitting is comparable to the $\rho(2150)-\rho(1900)$ splitting near $g=1.0$, lending support to the ``model (v)" scenario presented above.

With the isovector vector hybrid mass estimated to be 2100 MeV, the lattice and model results of Sect. \ref{sect:5} imply that an isoscalar vector hybrid should have a mass of 2100-2250 MeV, while the ``$s\bar s$" isoscalar should lie in the range 2220-2350 MeV.

\begin{figure}[ht]
\includegraphics[width=10cm,angle=0]{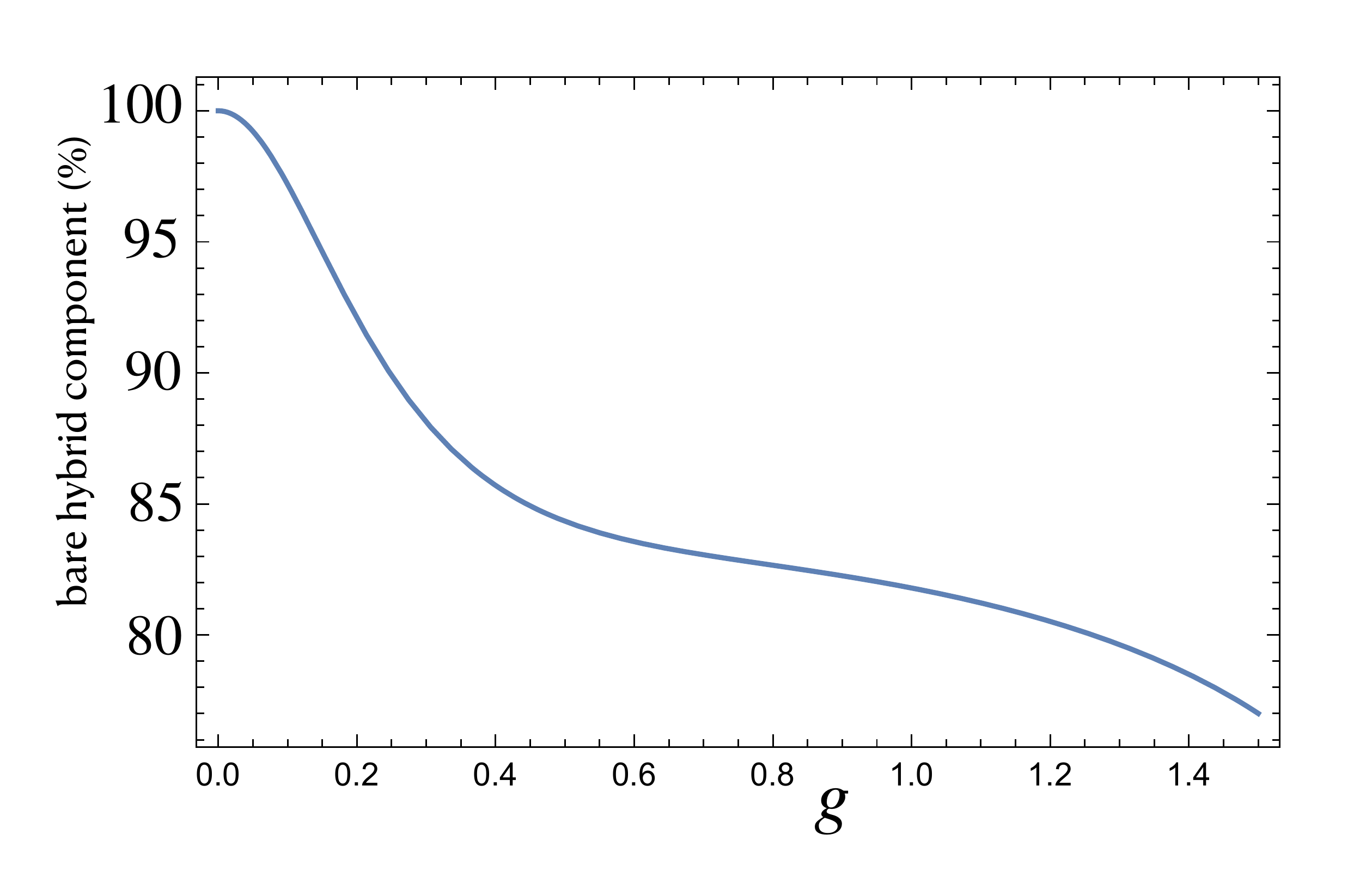}
\caption{$q\bar q g$ Fraction of Vector Hybrid.}
\label{fig:comp}
\end{figure}

If a vector hybrid is to be discovered in $e^+e^-$ annihilation its decay constant should be comparable to other excited $\rho$ states. Presumably this coupling is set by the $q\bar q$ content of the hybrid, which can be obtained from the mixing matrix just described. The bare $q \bar q$ component of the full hybrid is shown as a function of the strong coupling in Fig. \ref{fig:comp}, where it is seen that approximately 20\% of the state is $q\bar q$ in various configurations.

This observation can be used to compute the hybrid decay constant. Defining a decay constant in the usual way

\be
\langle 0 | \bar \psi \gamma^\mu \psi(0)|V(\lambda)\rangle = m_V f_V \epsilon^\mu(\lambda)
\ee
permits one to obtain the decay (neglecting electron masses)
\be
\Gamma(V\to e^+e^-) = \frac{4\pi \alpha^2 Q_{eff}^2}{3 m_V} f_V^2
\ee
where $Q_{eff} = Q_u/\sqrt{2} + Q_d/\sqrt{2}$ is the effective charge of an isovector meson in units of the electron charge. In the case of the hybrid meson, allowing the state to be a sum over  components gives

\be
f_H = \frac{1}{\sqrt{M_H}} \sum_{n \neq H} \sqrt{M_n} f_V^{(n)} C_n
\label{eq:fH}
\ee
where $C_n = \langle nS | H_1(1^{--})\rangle$ are the state components obtained previously.

Unfortunately, only the $\rho(770)$ decay constant is known, so we must model the remaining.
Following the methodology of Ref. \cite{Lakhina:2006vg}, the decay constant is written as

\begin{equation}
f_V^{(n)}=\sqrt{\frac{3}{M_n}}\int\frac{d^3k}{(2\pi)^3} \psi^{(n)}(\vec{k})
\sqrt{1+\frac{m_q}{E_k}} \sqrt{1+\frac{m_{\bar{q}}}{E_{\bar{k}}}}
\left(1+\frac{k^2}{3(E_k+m_q)(E_{\bar{k}}+m_{\bar{q}})}\right).
\label{eq:F}
\end{equation}
The mass in this expression originates in the relativistic normalization of the state vector, and is the reason the square roots of meson mass appear in Eq. \ref{eq:fH}.

Evaluating Eq. \ref{eq:F} for the $\rho(770)$ gives $f_\rho = 300$ MeV, to be compared to the experimental value of 220 MeV (the simple quark model is known to give decay constants that are too large). Other decay constants obtained in this way are $f_{\rho(2S)} = 160$ MeV, $f_{\rho(3S)} = 130$ MeV, $f_{\rho(4S)} = 110$ MeV, $f_{\rho(5S)} = 100$ MeV, and $f_{\rho(6S)} = 95$ MeV. Evaluating Eq. \ref{eq:fH} then yields

\be
f_{H_1(1^{--})} \approx 20 \ \textrm{MeV}.
\ee
where we have accounted for the tendency to over-predict decay constants. Thus one can expect production of the vector hybrid in $e^+e^-$ annihilation at approximately 1\% of the strength of canonical mesons.

\section{Conclusions}
\label{sect:8}

A constituent gluon model of gluodynamics has been explored. This model permits describing glueballs and hybrid mesons as simple bound states in a formalism that can be considered a minimal extension of typical constituent quark models. The model is commensurate with lattice field theory results and leverages the Hamiltonian of QCD in Coulomb gauge to describe the relevant dynamics.

This picture was used to model hybrid meson flavor mixing, assuming that hybrids and glueballs are dominated by the minimal number of quasigluons required by the state.
Hybrid flavor mixing is unique in that it can be described by low order diagrams because the quark-antiquark pair is in a color octet state. In this case annihilation can happen via a transverse gluon or an instantaneous (potential)  gluon. Both cases require quark annihilation in the ${}^{(2S+1)}L_J = {}^3S_1$ state, which in turn restricts substantial flavor mixing to the $H_1(0^{-+})$, $H_1(1^{-+})$, and $H_1(2^{-+})$ multiplets. Comparison to lattice results computed with a pion of mass 391 MeV show broad agreement, with the largest discrepancy in the vector multiplet. This discrepancy is difficult to reconcile in the present model, and if confirmed, likely implies that the quasigluon approximation needs to be abandoned or heavily modified. Of course, any new model must continue to explain the weak mixing computed in the $H_2$ (and presumably other) multiplets.  Alternatively, the model agrees broadly with lattice data, implying that the assumptions made may be reasonable, and that quasigluons do indeed serve as a useful description of low-lying gluonic excitations.

A similar computation of configuration mixing of hybrid and canonical mesons yielded reasonable agreement with an NRQCD computation for $J/\psi$ and $\Upsilon$ mixing (within large errors for both methods). Thus the quasigluon approach proves useful in this context as well.

Combining the model computations leads to a picture in which an isovector  vector hybrid is expected with a mass of approximately 2100 MeV. This vector does not mix extensively with canonical mesons, and has a decay constant of approximately 20 MeV, perhaps permitting its observation in $e^+e^-$ scattering. The state's  isospin partners are expected at 2100-2250 MeV and 2220-2350 MeV. Similarly, if the $\pi_1(1600)$ is confirmed then isospin partners are expected at 1750-1780 MeV and around 1900 MeV. This last state is a natural identification of the recently seen $\eta_1$.

The results of this investigation encourage further work on the hybrid spectrum with the goal of achieving detailed agreement. This will likely require the addition of higher order spin-dependent and spin-independent interactions. It will also be of interest to compute light hybrid strong decay rates as these will be crucial to identifying novel hadrons. Finally, the topic of radiative transitions of hybrid mesons remains relatively unexplored.

\acknowledgments

The author thanks Matthew Shepherd for rekindling his interest in light exotics and acknowledges support by the U.S. Department of Energy under contract DE-SC0019232.

\end{document}